# A wall function approach in lattice Boltzmann method: algorithm and validation using turbulent channel flow

Mengtao Han [a]*, Ryozo Ooka [a], Hideki Kikumoto [a]

[a] Institute of Industrial Science, The University of Tokyo, Tokyo, Japan

**Abstract (236)**
In the lattice Boltzmann method (LBM), the widely utilized wall boundary is the bounce-back (BB) boundary, which corresponds to the no-slip boundary. The BB boundary prevents the LBM from capturing the accurate shear drag on the wall when addressing high Reynolds number flows using coarse-grid systems. In this study, we proposed a "wall-function bounce (WFB)" boundary that incorporates a wall function into the LBM's boundary condition and overcomes the limitation of the BB. The WFB boundary calculates the appropriate shear drag on the wall using a wall function model, and thereafter modifies distribution functions to reflect the shear drag. The Spalding's law was utilized as the wall function in WFB. Simulations of turbulent channel flow at $\text{Re}_\tau = 640$ and 2003 using the LBM-based large-eddy simulation (LBM-LES) were conducted to validate the effectiveness of the proposed boundary condition. The results indicate that the BB boundary underestimated the time-averaged velocity in the buffer layer at $\text{Re}_\tau = 640$, and the averaged velocity in the entire domain at $\text{Re}_\tau = 2003$, when using coarse-grid systems. However, WFB obtained the proper shear drag on the wall and thus, compensated for the underestimation and agreed better with the experimental or DNS data, especially at the first-layer grid. In addition, WFB improved the Reynolds normal stress in the near-wall region to some extent. The distributions of shear stress on the wall by WFB was analogous to those by the wall model function in the finite volume method.

| Nomenclature | | | |
|---|---|---|---|
| BB | bounce-back boundary | $C_s$ | Smagorinsky constant |
| $D$ | half-height of the channel, m | $\mathbf{e}_a$ | discrete velocity vector of the virtual particle in the $a$-direction |
| $e_s$ | speed of sound in LBM | $F_a$ | external force term in the $a$-direction |
| $f_a$ | distribution function of the virtual particle | $N$ | number of total grids |
| $p$ | local pressure, Pa | $\mathbf{r}$ | position vector of the virtual particle |
| $\text{Re}_\tau$ | friction Reynolds number, $= u_\tau D/\nu$ | $T$ | parameter for time normalization, s, $T = D/u_\tau$ |
| $t$ | time step, s | $\mathbf{u}$ | velocity vector, m·s$^{-1}$ |
| $u_\tau$ | friction velocity, m·s$^{-1}$, $= \sqrt{\tau_w/\rho}$ | $u^+$ | dimensionless velocity, $= u/u_\tau$ |
| $\boldsymbol{\nu}$ | a virtual velocity vector utilized for the free-slip boundary, m·s$^{-1}$ | WFB | wall-function bounce boundary |
| $y^+$ | dimensionless distance from the wall, $= yu_\tau/\nu$ | $\Delta f$ | difference of the distribution function on wall boundary after and before WFB collision |
| $\Delta t$ | discrete time interval, s | $\epsilon_{u^+}$ | L2 error norm of $u^+$ |
| $\kappa$ | Karman constant | $\nu$ | molecular kinematic viscosity, m$^2$·s$^{-1}$ |
| $\rho$ | local fluid density, kg·m$^{-3}$ | $\boldsymbol{\tau}$ | stress, kg·m$^{-1}$·s$^{-2}$ |
| $\boldsymbol{\tau}_w$ | shear stress on the wall, kg·m$^{-1}$·s$^{-2}$ | $\Omega_a(\mathbf{r},t)$ | collision function |
| $x, y, z$ | streamwise, normal, and spanwise components of the spatial coordinate, respectively, m | $u_x, u_y, u_z$ | components of $\mathbf{u}$ in the $x$, $y$, and $z$ directions, respectively, m·s$^{-1}$ |
| $\nu_x, \nu_y, \nu_z$ | components of $\boldsymbol{\nu}$ in the $x$, $y$, and $z$ directions, respectively, m·s$^{-1}$ | $\sqrt{\langle u_x'^2 \rangle}, \sqrt{\langle u_y'^2 \rangle}, \sqrt{\langle u_z'^2 \rangle}$ | components of standard deviations of the fluctuating velocity in the $x$, $y$, and $z$ directions, respectively, m·s$^{-1}$ |
| Subscripts and Superscripts | | | |
| $f_a^*$ | distribution function $f_a$ updated by the collision step | $\widetilde{f}_a$ | distribution function $f_a$ that is modified by the WFB |
| $\Phi\vert_{(i,j,k)}$ | a property $\Phi$ or $\Phi_a$ at the grid $(i,j,k)$ | $\Phi\vert^t$ | a property $\Phi$ at time step $t$ |
| $\vert\boldsymbol{\Phi}\vert$ | magnitude of the vector property $\boldsymbol{\Phi}$ | $\langle\Phi\rangle$ | time-average of property $\Phi$ |
| $\overline{\Phi}$ | spatial-average of property $\Phi$ | | |

## 1. INTRODUCTION

In recent years, the lattice Boltzmann method (LBM) has been applied to high Reynolds number (high-Re) turbulent flows [1,2], such as the built environment [3–6], analogous to the conventional finite volume method (FVM). Based on the lattice Boltzmann equation (Equation 1), as well known, the LBM simulates the fluid motion by using distribution functions to represent the properties of the collection of particles. The simulation results are dependent on the collective collide-and-stream behavior of the particles in the system, rather than solving physical quantities on the macro scale [7]. In addition, the macroscopic fluid quantities are calculated through the integration of the distribution functions (Equation 2). Furthermore, the LBM can also be utilized for the large-eddy simulation (LBM-LES) to solve high-Re flows and several applications have been reported [1,3,4,8–12].

$$f_a(\mathbf{r} + \Delta t \mathbf{e}_a, t + \Delta t) - f_a(\mathbf{r}, t) = \Omega_a(\mathbf{r}, t) + F_a \qquad (1)$$

$$\rho = \sum_a f_a(\mathbf{r},t), \qquad \mathbf{u} = \frac{1}{\rho}\sum_a \mathbf{e}_a f_a(\mathbf{r},t), \qquad p = \rho e_s^2 \qquad (2)$$

In the viscous sublayer, which is a thin layer near the wall, the inertia force is in almost the same order as the viscous force, or only slightly larger than the viscous force, manifesting as Re~1. In this sublayer, the fluid suffers from viscosity such that the motion does not follow the free flow pattern. In other words, the fluid pattern in this sublayer should not be modeled with a high-Re model [13]. An ideal handling method is to densify the grids near the wall such that the fluid pattern in the sublayer can be calculated directly while not being modeled. Unfortunately, in many instances, the first-layer grid is not placed in the viscous sublayer when solving high-Re problems to reduce the computational cost, thus resulting in that unsatisfactory results in the near-wall region with the no-slip wall boundary. In the FVM, a widely utilized approach to overcome the deficiency is a wall function that models the appropriate velocity profile near the wall region. Some commonly used wall function models include the logarithmic law [14], the two-layer model [15], and the Spalding's law [16,17]. Among them, the Spalding's law is increasingly widely utilized in high-Re turbulence problems; because its curve is continuous and is the asymptote of the curves of both the viscous sublayer and the logarithmic layer, hence, it describes the law of the wall using one equation. Hitherto, the Spalding's law has been applied to the FVM [18–20].

In the LBM, the most widely adopted wall boundary is the bounce-back boundary condition [21,22], because of its advantages of the simple algorithm and easy implementation. In this boundary, after hitting the wall, the fluid particles completely bounce back to the path in which they come from rather than move forward. Therefore, no flux across the wall, and no relative transverse motion exist between the fluid and wall. Researchers have applied the bounce-back boundary to both isothermal and non-isothermal fluid problems [3,6,9,10,23], especially in relatively low-Re flow problems [24–26]. However, the bounce-back corresponds to the no-slip boundary, which may lead to a misprediction of the flow pattern in the viscous sublayer when solving high-Re flow problems. On the other hand, the widely adopted discrete velocity scheme in the LBM hitherto is the DdQq scheme [27] that typically utilizes uniform cubic lattices for the entire simulation domain. This may cause the misprediction of the shear drag on the wall by the bounce-back boundary more significant when using coarse-grid systems in solving high-Re flow. Currently, several new technologies are being developed to improve grid features such as the local grid refinement for the DdQq scheme [28,29], and the finite volume LBM with unstructured meshes[30]; however, these methods will not be discussed in this study.

In such situations, although the mass and momentum can be conserved, the bounce-back boundary cannot describe the shear drag on the wall accurately in turbulent flows, especially if the near-wall grids are coarse. Therefore, although the LBM-LES can be utilized in turbulent flows, it exhibits the same problem of mispredicting the shear drag on the wall as the FVM-LES. In some studies to solve relatively high-Re problems with LBM-LES, there are reports that the near-wall velocity errors may partly be due to defects of the bounce-back. Fernandino [2] indicated that the error of the results was apparent near the wall unless finer grid systems were employed in their simulation of the free surface flow in a wide rectangular duct using LBM-LES. Han reported that in the simulation of indoor flow [3] and the flow around a building [4] using LBM-LES, the simulated velocity near the walls agreed inaccurately with the experimental data; they supposed that it was partly owing to the lack of the wall function. It should be noted that the turbulent flow simulation using LBM-LES is challenging. It is because that several other factors will affect the result accuracy in addition to the boundary condition, such as the collision functions [31,32], SGS models, grid resolutions [33], and the number of discrete speeds of the lattice. However, in this study, we have mainly focused on the effect of the boundary conditions. Therefore, an enhanced wall treatment is necessary for LBM in addressing turbulent flows, similar to the wall function models in FVM.

Hitherto, reports regarding the progress of the wall function in LBM suited to turbulent flows are scarce. This is partly because the LBM is primarily applied to small-scale or not-too-high-Re flow problems; thus, the simulation accuracy is sufficient, and the implementation of the wall function is not urgent. Norouzi [34] implemented both the power-law and exponential wall functions into two relaxation time LBM (TRT-LBM) to consider the effect of the Knudsen layer in the transition flow regime. Their results of Poiseuille gas flow through a micro/nanochannel indicated that the TRT-LBM using the wall function could satisfactorily predict the flow behavior up to the upper end of the transition flow regime. Malaspinas and Sagaut [35] established a wall model for LBM-LES for high-Re wall-bounded flows, by relying on the analytical profile of the velocity profile within the first off-wall cell or the solution of turbulent boundary layer equations. Their model obtained accurate averaged velocity, Reynolds stress, and friction coefficient compared to the DNS or semi-analytical profiles in the turbulent channel flow at $Re_\tau$=1000, 2000, and 20000. Pasquali et al. [36] proposed a wall function model for the cumulant LBM that sets a partial slip velocity on the wall by computing a skin frictional coefficient. Their model yielded results that were in good agreement with the DNS data in the case of the velocity profile, Reynolds shear, and normal stresses for the turbulent channel flow at $Re_\tau$=950, 2000, and 16000.

In this study, we proposed a new boundary condition, the "wall-function bounce (WFB)" boundary for walls, which incorporates a wall function model into the LBM's boundary condition so that the LBM can be applied for turbulent flows. The Spalding's law was employed for the incorporation, because of its prevalence in the FVM when solving turbulent flows. LBM-LES simulations of the turbulent channel flow at $Re_\tau = 640$ and $2003$ were employed to validate the effectiveness of the WFB boundary. The turbulent channel flow is a high-Re flow problem and is sensitive to the drag on the wall, which makes it appropriate for validating the effectiveness of the WFB boundary.

## 2. Algorithm of wall-function bounce (WFB) boundary for LBM
### 2.1. How can a wall function model be realized in LBM

For brevity, we hereinafter abbreviate the bounce-back boundary and proposed wall-function bounce boundary as BB and WFB, respectively. We first compare the difference between the no-slip (BB) boundary and the free-slip boundary in the LBM to obtain the principal idea of the realization of a wall function model. These explanations are based on the two-dimensional D2Q9 scheme [27] (see Appendix: Fig.A-1a) for clarity. Subsequently, we derive the WFB in detail based on the three-dimensional D3Q19 scheme (see Appendix: Fig.A-1b).

In the LBM, the most straightforward approach for the walls is the BB boundary, which corresponds to the no-slip wall boundary. In this boundary, the particles hit a wall while not penetrating it, implying that no particles move across the boundary. Subsequently, the particles bounce back rather than bounce forward, implying that no relative transverse movement occurs between the fluid and boundary (i.e., the fluid velocity on the wall is zero). Take the D2Q9 discrete velocity as an example (see Appendix: Fig.A-2), this process can be described as [37]:

$$f_2^*|_{(i,0)} = f_4|_{(i,0)}, \quad f_6^*|_{(i,0)} = f_8|_{(i,0)}, \quad f_5^*|_{(i,0)} = f_7|_{(i,0)} \tag{3}$$

Here, $f_a^*$ means $f_a$ updated by the collision step. With increasing Re, the uniform grid system near the walls becomes too coarse to capture the accurate shear drag at the wall and will consequently affect the accuracy of the subsequent calculations. Hence, it is necessary to introduce a new boundary that contains a wall function to overcome this shortcoming of the BB boundary.

Meanwhile, in the free-slip boundary, the distribution functions on the boundary become mirror-symmetrical about the boundary after the collision step (see Appendix: Fig.A-3), as Equation (4). Therefore, the normal velocity at the boundary grids toward the boundary is zero. However, the tangential velocity at the boundary remains.

$$f_2^*|_{(i,0)} = f_4|_{(i,0)}, \quad f_6^*|_{(i,0)} = f_7|_{(i,0)}, \quad f_5^*|_{(i,0)} = f_8|_{(i,0)} \tag{4}$$

By comparing Equation (3) and (4), it is clear that in the stream process, a particle moves to the boundary grid; subsequently, in the collision step, the particle bounces back to the incident direction in the no-slip boundary (Fig. 1a) or bounces forward in the free-slip boundary (Fig. 1b). On the other hand, if a wall function is implemented, its effect is equivalent to applying a reverse drag force on the velocity direction to the wall compared to the free-slip boundary, causing the particles near the wall to decelerate, as demonstrated in Fig. 1c.

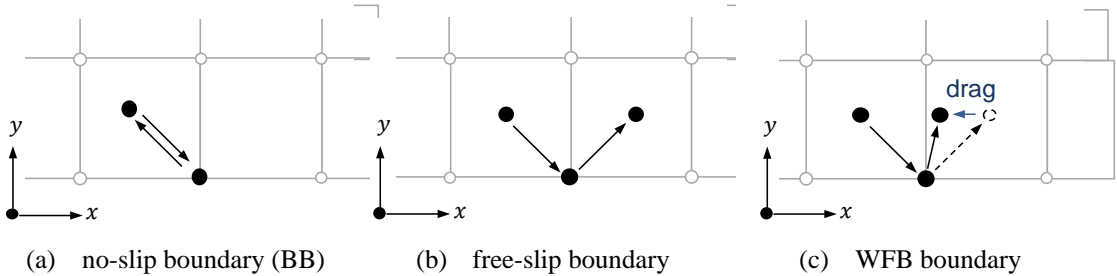

(a) no-slip boundary (BB)    (b) free-slip boundary    (c) WFB boundary

Fig. 1 The difference among no-slip, free-slip, and WFB boundaries

We observed that the difference in the distribution functions between the BB and free-slip boundaries is just that the values of $f_5^*$ and $f_6^*$ (i.e., distribution functions in the diagonal directions and pointing to the interior of the flow field) are swapped. Therefore, it is straightforward that we can reassign the values of $f_5^*$ and $f_6^*$, to achieve the effect of the wall function (drag).

### 2.2. Derivation of WFB boundary

We now return to the three-dimensional problems and derive the algorithm of WFB for D3Q19 scheme. The core task is to determine the adjustment of the distribution functions on the boundary, which are in the diagonal directions and pointing to the interior of the flow field (analogy with $f_5^*$ and $f_6^*$ in D2Q9 scheme). We assume that the plane normal to the *y*-direction is the wall boundary and the positive direction of *y* points to the flow field. In addition, positive *x* is the streamwise direction. To clarify, from here to the end of this section, we define that $\widetilde{f_a}$ is the distribution function $f_a$ revised by the wall function, and $f_a^*$ represents $f_a$ updated by the collision step. $\rho$ and **u** are the density and velocity obtained by the WFB boundary, respectively.

According to the boundary layer theory, the momentum equation on the boundary can be simplified as Equation (5). In the near-wall region, the flow pattern can be regarded as a simple two-dimensional shear flow parallel to the wall, and the streamwise is positive *x*-direction. Equation (5) indicates that the variation of the momentum $\rho \mathbf{u}$ of this shear flow is resulted by the shear stress $\boldsymbol{\tau}$.

$$\frac{\partial \rho \mathbf{u}}{\partial t} = -\frac{\partial \boldsymbol{\tau}}{\partial y} + \cdots \tag{5}$$

Here, $\cdots$ is other terms such as pressure term, which we do not focus on in this study.

As shown in Fig. 2, by discretizing Equation (5), we have a general form as follows.

$$\frac{\rho \mathbf{u}|_{(i,1,k)}^{t} - \rho \mathbf{u}|_{(i,1,k)}^{t-\Delta t}}{\Delta t} = -\frac{\boldsymbol{\tau}|_{(i,1+\frac{1}{2},k)} - \boldsymbol{\tau}|_{(i,1-\frac{1}{2},k)}}{\Delta y} + \cdots \tag{6}$$

Equation (6) can be decomposed into Equation (7) and (8).

$$\frac{\rho \boldsymbol{v}|_{(i,1,k)}^{t} - \rho \mathbf{u}|_{(i,1,k)}^{t-\Delta t}}{\Delta t} = -\frac{\boldsymbol{\tau}|_{(i,1+\frac{1}{2},k)} - 0}{\Delta y} + \cdots \tag{7}$$

$$\frac{\rho \mathbf{u}|_{(i,1,k)}^{t} - \rho \boldsymbol{v}|_{(i,1,k)}^{t}}{\Delta t} = -\frac{0 - \boldsymbol{\tau}|_{(i,1-\frac{1}{2},k)}}{\Delta y} = \frac{\boldsymbol{\tau}_w|_{(i,0,k)}}{\Delta y} \tag{8}$$

Here we introduce a virtual momentum $\rho \boldsymbol{v}$, which is the momentum in the condition of the free-slip boundary (i.e., no shear drag on the boundary). This is a hypothetical intermediate status for the momentum variation between time step $t - \Delta t$ and $t$.

In Equation (7), the stream process changes $\rho \mathbf{u}|_{(i,1,k)}^{t-\Delta t}$ to the virtual intermediate status $\rho \boldsymbol{v}|_{(i,1,k)}^{t}$ in the condition of the free-slip boundary we hypothesize. The momentum variation is resulted by $\boldsymbol{\tau}|_{(i,1+\frac{1}{2},k)}$ while $\boldsymbol{\tau}|_{(i,1-\frac{1}{2},k)} = 0$ because there is no shear drag on the free-slip boundary. This process implemented via the stream step in the LBM and does not require any special treatment.

Next, in Equation (8), the shear drag $\boldsymbol{\tau}_w|_{(i,0,k)}$ on the wall is introduced into the collision step, thus $\rho \boldsymbol{v}|_{(i,1,k)}^{t}$ turns to $\rho \mathbf{u}|_{(i,1,k)}^{t}$. This process is implemented via the collision step and realized by WFB. And $\rho \mathbf{u}|_{(i,1,k)}^{t}$ is the correct momentum that we obtain.

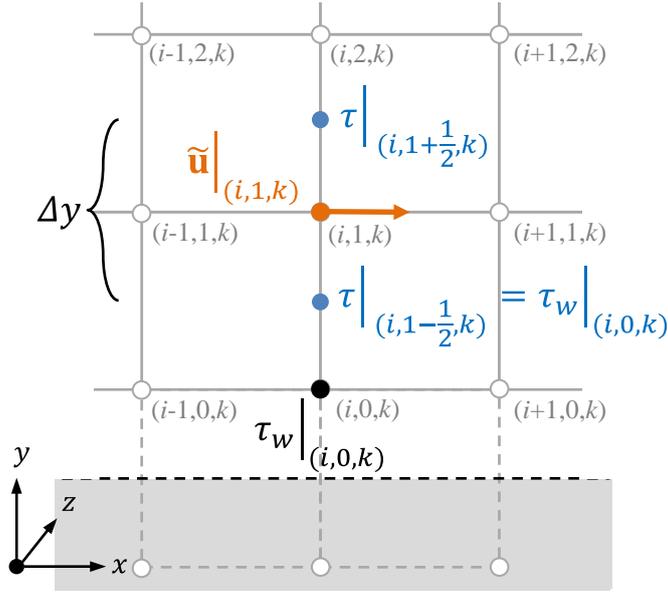

Fig. 2 Relationship between momentum variation and shear drag. The shear drag on the wall causes variations in the momentum.

Next, we can decompose Equation (8) along *x*- and *z*-directions (*y*-direction is the normal direction and the corresponding velocity component is zero). Thus, $\boldsymbol{\tau}_w$ can be decomposed to *x*-component $\tau_{w,x}$ and *z*-component $\tau_{w,z}$. Likewise, $\rho \mathbf{u}$ and $\rho \boldsymbol{v}$ are decomposed to $\rho u_x$, $\rho u_z$, and $\rho v_x$, $\rho v_z$, respectively. Therefore, we have

$$\rho u_x|_{(i,1,k)} - \rho v_x|_{(i,1,k)} = \frac{\Delta t}{\Delta y} \tau_{w,x}|_{(i,0,k)} \tag{9a}$$

$$\rho u_z|_{(i,1,k)} - \rho v_z|_{(i,1,k)} = \frac{\Delta t}{\Delta y} \tau_{w,z}|_{(i,0,k)} \tag{9b}$$

The above equation indicates that in the *x*-component, there is a momentum loss ($\frac{\Delta t}{\Delta y} \tau_{w,x}$) caused by the shear drag $\tau_{w,x}$ compared to the free-slip boundary, resulting in that the momentum near the wall reduces from $\rho v_x$ of the hypothetical free-slip boundary to $\rho u_x$ of the actual WFB boundary. The same situation also happens in the *z*-component.

The next problem is to convert the momentum to the distribution functions. By considering the particle velocity of D3Q19 (see Appendix: Table A-1), we can expand Equation (2) on the free-slip boundary as

$$f_0 + f_1 + f_2 + f_3 + f_4 + f_5 + f_6 + f_7 + f_8 + f_9 + f_{10} + f_{11} + f_{12} + f_{13} + f_{14}$$
$$+ f_{15} + f_{16} + f_{17} + f_{18} = \rho \tag{10a}$$

$$f_1 - f_2 + f_7 - f_8 + f_9 - f_{10} + f_{13} - f_{14} + f_{15} - f_{16} = \rho v_x \tag{10b}$$

$$f_3 - f_4 + f_7 - f_8 + f_{11} - f_{12} - f_{13} + f_{14} + f_{17} - f_{18} = \rho v_y = 0 \tag{10c}$$

$$f_5 - f_6 + f_9 - f_{10} + f_{11} - f_{12} - f_{15} + f_{16} - f_{17} + f_{18} = \rho v_z \tag{10d}$$

Meanwhile, in the condition of WFB boundary, the density $\rho$ conserves while the velocity varies to $u_x, u_y$, and $u_z$. Here, $v_y = u_y = 0$ is tenable. Based on our previous assumptions, only the distribution functions in the diagonal direction and pointing to the interior of the flow field are reassigned while those in the direction of the axis remain unchanged. Therefore, $f_7$, $f_{11}, f_{14}$, and $f_{17}$ become $\widetilde{f_7}$, $\widetilde{f_{11}}, \widetilde{f_{14}}$, and $\widetilde{f_{17}}$ in WFB boundary, respectively. This means, in the WFB, we have

$$f_0 + f_1 + f_2 + f_3 + f_4 + f_5 + f_6 + \widetilde{f_7} + f_8 + f_9 + f_{10} + \widetilde{f_{11}} + f_{12} + f_{13} + \widetilde{f_{14}}$$
$$+ f_{15} + f_{16} + \widetilde{f_{17}} + f_{18} = \rho \tag{11a}$$

$$f_1 - f_2 + \widetilde{f_7} - f_8 + f_9 - f_{10} + f_{13} - \widetilde{f_{14}} + f_{15} - f_{16} = \rho u_x \tag{11b}$$

$$f_3 - f_4 + \widetilde{f_7} - f_8 + \widetilde{f_{11}} - f_{12} - f_{13} + \widetilde{f_{14}} + \widetilde{f_{17}} - f_{18} = \rho u_y = 0 \tag{11c}$$

$$f_5 - f_6 + f_9 - f_{10} + \widetilde{f_{11}} - f_{12} - f_{15} + f_{16} - \widetilde{f_{17}} + f_{18} = \rho u_z \tag{11d}$$

In addition, we assume that $f_7, f_{14}$ are independent of $f_{11}, f_{17}$. Therefore, according to Equation (9)–(11), we have

$$\widetilde{f_7} + \widetilde{f_{14}} = f_7 + f_{14}, \quad \widetilde{f_{11}} + \widetilde{f_{17}} = f_{11} + f_{17} \tag{12a}$$

$$\widetilde{f_7} - \widetilde{f_{14}} - f_7 + f_{14} = \frac{\Delta t}{\Delta y}\tau_{w,x}, \quad \widetilde{f_{11}} - \widetilde{f_{17}} - f_{11} + f_{17} = \frac{\Delta t}{\Delta y}\tau_{w,z} \tag{12b}$$

Then $\widetilde{f_7}, \widetilde{f_{11}}, \widetilde{f_{14}}, \widetilde{f_{17}}$ can be solved as

$$\widetilde{f_7} = f_7 + \frac{\Delta t}{2\Delta y}\tau_{w,x}, \quad \widetilde{f_{14}} = f_{14} - \frac{\Delta t}{2\Delta y}\tau_{w,x} \tag{13a}$$

$$\widetilde{f_{11}} = f_{11} + \frac{\Delta t}{2\Delta y}\tau_{w,z}, \quad \widetilde{f_{17}} = f_{17} - \frac{\Delta t}{2\Delta y}\tau_{w,z}, \tag{13b}$$

We notice that $f_7, f_{11}, f_{14}, f_{17}$ in Equation (13) are after the collision under the free-slip boundary. Before the collision step, they correspond to $f_{13}, f_{18}, f_8, f_{12}$, respectively. Therefore, we can conclude the integrated collision step of WFB boundary as

Han et al.

$$\widetilde{f_7}^* = f_{13} + \frac{\Delta t}{2\Delta y}\tau_{w,x}, \quad \widetilde{f_{14}}^* = f_8 - \frac{\Delta t}{2\Delta y}\tau_{w,x} \tag{14a}$$

$$\widetilde{f_{11}}^* = f_{18} + \frac{\Delta t}{2\Delta y}\tau_{w,z}, \quad \widetilde{f_{17}}^* = f_{12} - \frac{\Delta t}{2\Delta y}\tau_{w,z} \tag{14b}$$

$$\begin{aligned}&f_1^* = f_2, \; f_2^* = f_1, \; f_3^* = f_4, \; f_4^* = f_3, \; f_5^* = f_6, \; f_6^* = f_5, \\ &f_8^* = \widetilde{f_7}^*, \; f_9^* = f_{10}, \; f_{10}^* = f_9, \; f_{12}^* = \widetilde{f_{11}}^*, \; f_{13}^* = \widetilde{f_{14}}^*, \\ &f_{15}^* = f_{16}, \; f_{16}^* = f_{15}, \; f_{18}^* = \widetilde{f_{17}}^*.\end{aligned} \tag{14c}$$

Equation set (14) is the algorithm of the WFB. The WFB only occurs on the wall boundary grids and mainly reflects in the collision step, while the stream step is unchanged. Equations (14a) (14b) are the core operations that implement the wall function. They indicate that the near-wall layer grids acquire an additional shear velocity, which has been reduced by the shear drag $\boldsymbol{\tau}_w$ on the wall. From equation (14) we can see that the sum of distribution functions before and after the WFB collision step is equal (i.e., sum of $f_a$ is equal to that of $f_a^*$), further indicating the conservation of mass in the WFB boundary, and it will also be confirmed in the following case study. $\tau_{w,x}$ and $\tau_{w,z}$ are the $x$- and $z$-components of $\boldsymbol{\tau}_w$ obtained from the wall function, respectively, and are solved as Equation (15). The negative sign indicates that the shear drag $\tau_{w,x}$ and $\tau_{w,z}$ should be in the reverse direction of the corresponding components $u_x$ and $u_z$ of the shear velocity.

$$\tau_{w,x} = -\frac{u_x}{\sqrt{u_x^2 + u_z^2}}\tau_w, \quad \tau_{w,z} = -\frac{u_z}{\sqrt{u_x^2 + u_z^2}}\tau_w \tag{15}$$

### 2.3. Calculation of shear drag $\tau_w$: Spalding's law

The remaining task is to obtain the correct shear drag $\tau_w$ on the wall. In this study, we choose the Spalding's law [16,17] as the wall function to be incorporated into the boundary, as shown in Equation (16).

$$y^+ = u^+ + e^{-\kappa \cdot B}\left[e^{\kappa \cdot u^+} - 1 - (\kappa \cdot u^+) - \frac{(\kappa \cdot u^+)^2}{2} - \frac{(\kappa \cdot u^+)^3}{6}\right] \tag{16}$$

Here, $\kappa = 0.41$ and $B = 5.5$. $y^+$ and $u^+$ are the distances from the wall and velocity nondimensionalized by the friction velocity $u_\tau$, as

$$y^+ = \frac{yu_\tau}{\nu}, \quad u^+ = \frac{|\mathbf{u}|}{u_\tau}, \quad u_\tau = \sqrt{\frac{\tau_w}{\rho}} \tag{17}$$

Since the Spalding's law is an implicit equation, we need an initial value and a root-finding algorithm (here, we have utilized the Newton's method). In a single step of WFB collision, the initial $u_\tau$ is first calculated using the distance of the first-layer grid and the shear velocity on it. Thereafter, it is substituted into Equation (16), and more approximated values of $u_\tau$ and $\tau_w$ are obtained after several iterations of the Newton's method. Lastly, the corrected value of $\tau_w$ is reflected in the distribution functions using Equations (14) and (15). After completing these operations, we shift to the next stream step.

So far, the WFB boundary has been completely implemented, and two important points should be explained here. One is that the WFB boundary is not restricted to the Spalding's law that was utilized in this study. On the contrary, it is a frame that can incorporate any type of wall function model into the LBM, given that the candidate model is capable of providing the appropriate shear drag $\tau_w$.

The other point is that although WFB is achieved based on the standard BB boundary, we believe that this concept has the potential to be extended to a more complex BB boundary. For example, in the Bouzidi-Firdaouss-Lallemand linear interpolated BB scheme [38], the last fluid grid near the wall was the interpolation function of this grid and the adjacent fluid grid in the last time step, and it was related to the dimensionless distance, $q$, of the fluid grid to the wall. Therefore, it is not difficult to add the shear drag, $\tau_w$, into the interpolation. The essential point to note is that $\tau_w$, $\Delta t$, and $\Delta y$ should be determined using the dimensionless distance, $q$.

## 3. Simulation and validation
### 3.1. Simulated case description

Turbulent channel flow was employed to evaluate the proposed boundary. The simulation domain depicted in Fig. 3, is the same as that proposed by Moin and Kim [39]; the lengths along the streamwise direction $x$, normal direction $y$, and span direction $z$ components were $2\pi D$, $2D$, and $\pi D$, respectively. $\text{Re}_\tau = 640$ and $2003$ were utilized for the validation ($\text{Re}_\tau = u_\tau D/\nu$, $D$: half height of channel; $u_\tau$: the friction velocity). Simulations of LBM-LES were implemented using the MRT model [40]. To simplify the problem, we chose the standard Smagorinsky SGS model ($C_s = 0.1$ [41]) and added the van-Driest style damping function [42]. The *x*- and *z*-direction boundaries were periodic, and the *y*-direction boundary was the wall where we implemented the BB or the WFB boundary.

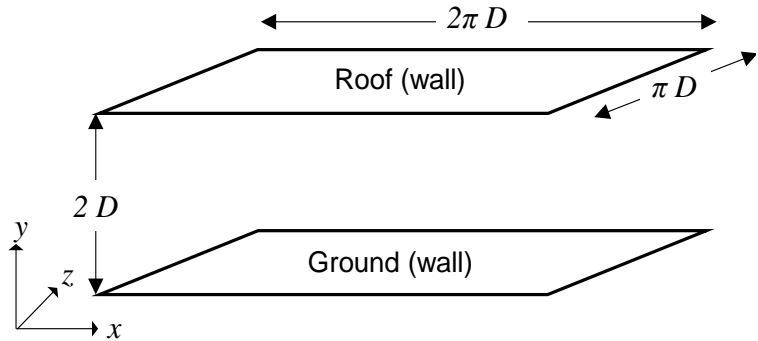

Fig. 3 Simulation domain of the turbulent channel flow. Top and bottom are infinite planes, and all other sides are periodic boundary conditions. The height of the channel is $2D$, with $D = 0.1\ m$. The length and width are $2\pi D$ and $\pi D$, respectively.

The case settings are listed in Table 1. To validate the wall function, the first-layer grid should be in the logarithmic layer or the buffer layer; i.e., the $y^+$ of the first-layer grid should be larger. However, it is difficult to place the first-layer grid in the logarithmic layer because it may cause the grid system to become too coarse and the results unacceptable. In this study, we set the grid resolutions according to the principle of placing the first-layer grid in the buffer layer ($\sim 10 < y^+ <$

Han et al.

~100). Although these grid resolutions may be coarse to obtain sufficient satisfying results throughout the channel, they are suitable for validating the wall function. Because the LBM is a weakly-compressible method for addressing incompressible flows, the compressibility errors may occur if an improper time interval is utilized [43–45]. Therefore, according to our pre-tests, the time intervals of each case were set to be considerably small to make sure that the $\text{Ma}_{LB} \ll 0.3$ and to avoid apparent compressibility errors [44,45].

**Table 1** Case settings.

| Test case | $\text{Re}_\tau$ and Re* | Grid resolutions | Mesh size ($x \times y \times z$) | $y^+$ at the first layer grid | Time interval (s) | $\text{Ma}_{LB}$ | y-direction boundary condition |
|---|---|---|---|---|---|---|---|
| A20_BB | $\text{Re}_\tau = 640$; Re~13800 | $D/20$ | $128 \times 40 \times 64$ | 32 | 1/800 | 0.060 | BB |
| A20_WFB | | | | | | | WFB |
| A40_BB | | $D/40$ | $256 \times 80 \times 128$ | 16 | 1/1600 | | BB |
| A40_WFB | | | | | | | WFB |
| B40_BB | $\text{Re}_\tau = 2003$; Re~48000 | $D/40$ | $256 \times 80 \times 128$ | 50 | 1/4800 | 0.069 | BB |
| B40_WFB | | | | | | | WFB |
| B80_BB | | $D/80$ | $512 \times 160 \times 256$ | 25 | 1/9600 | | BB |
| B80_WFB | | | | | | | WFB |

*: Re is defined by the time-averaged velocity in the middle of the channel.

$\langle u \rangle$ of the first-layer grid near the wall and the middle-layer grid was monitored for all the BB cases to ensure the time convergence (Fig. 4). $\langle u \rangle$ became stable after approximately $20T$ ($T = D/u_\tau$), and the data at $92T$ were utilized as they reached the time convergence. Thereafter, the time-averaged values of the same layer grids in the simulation domain were spatially averaged and utilized as the final result of every y-coordinate. After the sufficient sampling period, the force balance state was achieved and $u_\tau$ generally satisfied the theoretical value. Therefore, the results were normalized using $D$ and theoretical value of $u_\tau$.

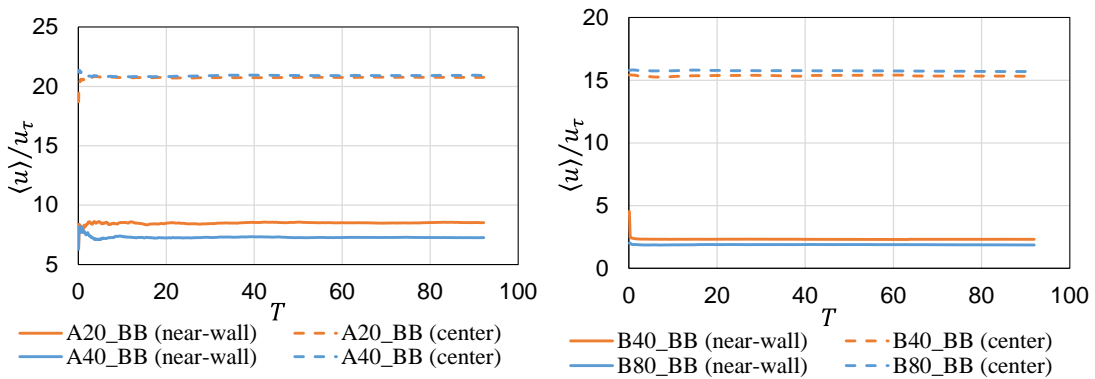

Fig. 4 Normalized values $\langle u \rangle$ of the first-layer grid near the wall and the middle-layer grid at two different frictional $\text{Re}_\tau = 640$ (left), and $\text{Re}_\tau = 2003$ (right). All the values became stable after approximately $20T$; data at $92T$ were utilized as they reached the time convergence.

## 3.2. Results and discussions
### 3.2.1. Confirmation of mass conservation in WFB

The mass conservation in the WFB had to be first confirmed. Here, we examined the differences between the distribution functions $f_a$ before the WFB collision and $\widetilde{f_a}^*$ after the WFB collision at all the boundary grids. 5000 simulation steps were checked in all the WFB cases when the turbulence was fully developed. Appendix B shows the detailed results.

We found that approximately 11–14 % samples of the differences were not zero. However, the orders of the differences were in the range of approximately $1/10^{-16} - 1/10^{-14}$ of the mean $f_a$ at the wall boundary, further indicating that the non-zero differences were considerably smaller than $f_a$, and that it was probably caused by the precision errors of the computer while calculating the floating-point numbers. Therefore, the "input" distribution functions (or mass) before the WFB collision were generally equal to the "output" after the collision, which demonstrates the conservation of mass in the WFB.

### 3.2.2. Results of time-averaged velocity

The time-averaged velocity profiles are depicted in Fig. 5. The experimental data from Hussain and Reynolds [46], and Clark [47] were used as the basis for $Re_\tau = 640$; the DNS data from Hoyas and Jiménez [48] were used for $Re_\tau = 2003$.

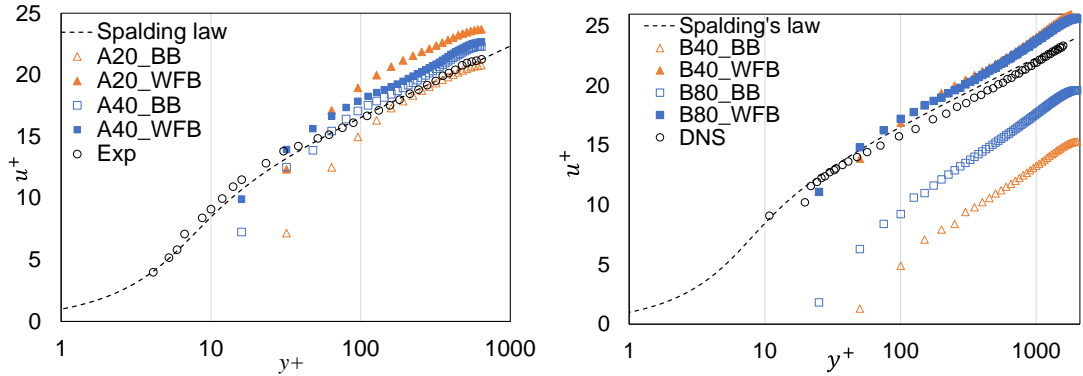

Fig. 5 Normalized velocity profiles for two different grid resolutions with two different wall boundary conditions in the condition of $Re_\tau = 640$ (left), and $Re_\tau = 2003$ (right).

At $Re_\tau = 640$, an underestimation of the time-averaged velocity was clearly observed in A20_BB and A40_BB in the buffer layer ($y^+ < \sim 100$), especially at the first-layer grid. With the increase in grid resolution, the deviations were compensated partially. Meanwhile, in A20_WFB and A40_WFB, the first-layer grids were generally on the Spalding's law line and they agreed with the experimental data better than the BB cases. However, in the logarithmic layer ($y^+ \geq \sim 100$), A20_WFB overestimated the velocity whereas A20_BB agreed better with the DNS data.

At $Re_\tau = 2003$, the differences between BB and WFB became significantly larger. Both B40_BB and B80_BB underestimated $u^+$ in both buffer and logarithmic layers. This evidently shows that the BB boundary mispredicted $\tau_w$, which consequently affected the accuracy of the time-averaged velocity. While utilizing WFB, the underestimation of $u^+$ was compensated, and both B40_WFB and B80_WFB agreed better with the DNS data. In particular, both the cases coincided with the Spalding's law in the buffer layer, although discrepancies were still visible in the logarithmic layer

Han et al.

to some extent.

The error analysis was conducted by utilizing the L2 error norm $\epsilon_{u^+}$ [49] defined by Equation (19), and the Spalding's law was utilized as the basis. $u^+_{LBM(i)}$ and $u^+_{Spalding(i)}$ represented $u^+$ at $y^+ = i$ and were obtained using the LBM and Spalding's law, respectively. A smaller $\epsilon_{u^+}$ demonstrated a smaller deviation between the simulation and the Spalding's law, and thus, exhibited a higher accuracy. The errors in the buffer layer, logarithmic layer, and the entire domain (including both buffer and logarithmic layer) were examined and are listed in Table 2, respectively.

$$\epsilon_{u^+} = \sqrt{\frac{\sum_i \left(u^+_{LBM(i)} - u^+_{Spalding(i)}\right)^2}{\sum_i {u^+_{Spalding(i)}}^2}} \tag{19}$$

Table 2 L2 error norm $\epsilon_{u^+}$ of all cases in the buffer layer, logarithmic layer, and the entire domain.

| $Re_\tau$ | 640 | | | | 2003 | | | |
|---|---|---|---|---|---|---|---|---|
| Case name | A20_BB | A20_WFB | A40_BB | A40_WFB | B40_BB | B40_WFB | B80_BB | B80_WFB |
| buffer layer ($y^+ < \sim 100$) | 0.183 | 0.069 | 0.103 | 0.071 | 0.911 | 0.049 | 0.617 | 0.057 |
| logarithmic layer ($y^+ \geq \sim 100$) | 0.009 | 0.055 | 0.039 | 0.067 | 0.420 | 0.080 | 0.224 | 0.066 |
| entire domain | 0.056 | 0.057 | 0.048 | 0.067 | 0.429 | 0.079 | 0.235 | 0.065 |

In all BB cases, the error of the buffer layer was larger than that of the logarithmic layer, and this affected the accuracy of the entire domain. Furthermore, this error became larger with the increase in $Re_\tau$; however, it became smaller with the increase in grid resolution. This also demonstrates that BB boundary will reduce the near-wall accuracy in simulating high-Re flows when using coarse grid systems. While using WFB, the accuracy of the buffer layer was observed to improve, and the errors became stable (approximately 5 %) in all the cases. This indicates that the WFB boundary implemented a proper shear drag on the wall by following the Spalding's law, and corrected the near-wall velocity.

As stated previously, the velocity overestimation in the logarithmic layer occurred in A20_WFB, thereby resulting in a larger $\epsilon_{u^+}$ when compared to A20_BB; however, the reason is not apparent yet. As discussed in the introduction, the accuracy of the results in the off-wall region was comprehensively affected by several other factors in addition to the wall boundary, such as SGS models, collision functions, and discrete velocity schemes. In this study, the effect of the wall boundary on the logarithmic layer was probably not as significant as other factors in the not-too-high-Re ($Re_\tau = 640$) case. Meanwhile, the averaged velocity of WFB in the logarithmic layer at $Re_\tau = 2003$ was better than that of BB, demonstrating that the wall boundary influenced the accuracy of the off-wall region greater in the higher-Re case.

In the logarithmic layer ($y^+ \geq \sim 100$), the inertia force became predominant, and all the cases followed the logarithmic law expressed in the following equation:

$$\frac{|\langle \mathbf{u} \rangle|}{u_\tau} = \frac{1}{\kappa} \ln y^+ + b \tag{20}$$

Here, $\kappa = 0.41$ is the Karman constant, whereas the value of $b$ refers to the prevalent range of 4.8–5.9 with respect to the experiment based on different conditions [50]. In this study, the value of $b$ for all cases are listed in Table 3. The time-averaged velocity in the logarithmic layer for all the cases was overestimated to some extent compared to the experimental or DNS data. The reason for this may be complex and can be attributed to factors such as grid resolutions, collision functions, and so on. However, all the cases were complied with the logarithmic law in general. Notably, negative values occurred in B40_BB and B80_ BB, which further stated that the velocity was universally underestimated due to the improper shear drag produced by the BB boundary.

Table 3 Value of $b$ for each case.

| Case name | $b$ | Case name | $b$ |
|---|---|---|---|
| A20_BB | 5.6 | B40_BB | -0.1 |
| D20_WFB | 6.4 | B40_WFB | 6.6 |
| D40_WFB | 6.6 | B80_ BB | -4.1 |
| A40_BB | 6.2 | B80_WFB | 6.5 |
| Exp [46] | 5.4 | DNS [48] | 5.4 |

### 3.2.3. Results of Reynolds stress

Fig. 6 shows the Reynolds normal and shear stresses for all the cases, respectively. At $Re_\tau = 640$, the experimental data proposed by Clark [47] were added to validate $\sqrt{\langle u_y'^2 \rangle}$ and $\sqrt{\langle u_z'^2 \rangle}$ in addition to the data proposed by Hussain and Reynolds [46] for validating $\sqrt{\langle u_x'^2 \rangle}$. It was observed that WFB improved the normal stresses near the wall to varying degrees at both $Re_\tau = 640$ and 2003 when compared to BB cases. In particular, for $\sqrt{\langle u_y'^2 \rangle}$ and $\sqrt{\langle u_z'^2 \rangle}$ at $Re_\tau = 2003$, the variation was larger than that at $Re_\tau = 640$. This suggests that the effect of WFB on the wall stress is more obvious to some extent in the higher-Re flow simulation. However, $\sqrt{\langle u_y'^2 \rangle}$ was underestimated at $y^+ < \sim 100$ by both the BB and WFB cases at $Re_\tau = 640$, which was also reported by Piomelli [51]. They suggested that this was a typical tendency in the close-to-wall region if the grids are not sufficiently fine. Meanwhile, $\sqrt{\langle u_z'^2 \rangle}$ at $Re_\tau = 640$ was overestimated near the channel center in all the BB and WFB cases. In addition, there were also deviations in $\sqrt{\langle u_x'^2 \rangle}$ and $\sqrt{\langle u_y'^2 \rangle}$ approximately at $y^+ = 400$ at $Re_\tau = 2003$ in both BB and WFB cases. A possible reason for these deviations was the coarse grid systems; however, the collision functions or discrete velocity scheme may also have affected the results, as stated earlier. For the shear stress, shear-stress profiles obtained by WFB attained the equilibrium shape that balanced the pressure gradient in the regions away from the walls, similar to that obtained by BB. In the near-wall region, Reynolds shear stress and viscous shear stress together balanced the pressure gradient. At the first-layer grid, Reynolds shear stress reproduced by BB was closely related to the grid resolution. WFB

Han et al.

improved the stress and reduced the effect of the grid resolution, so that the value became closer.

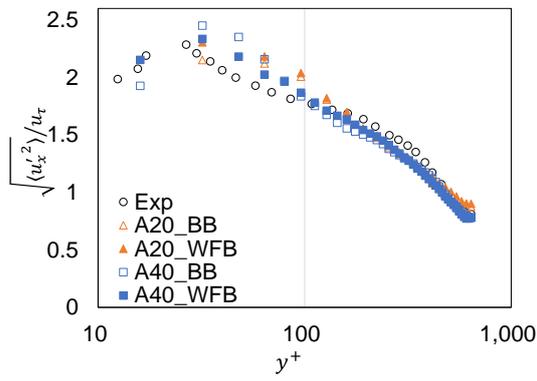

(a) $\sqrt{\langle u_x'^2 \rangle}$, $\mathrm{Re}_\tau = 640$

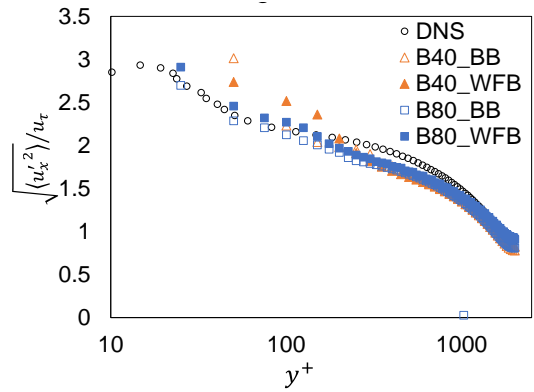

(b) $\sqrt{\langle u_x'^2 \rangle}$, $\mathrm{Re}_\tau = 2003$

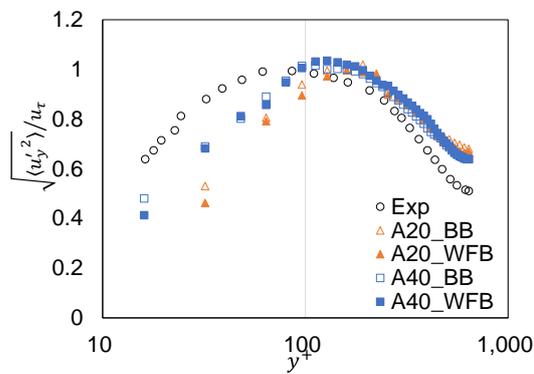

(c) $\sqrt{\langle u_y'^2 \rangle}$, $\mathrm{Re}_\tau = 640$

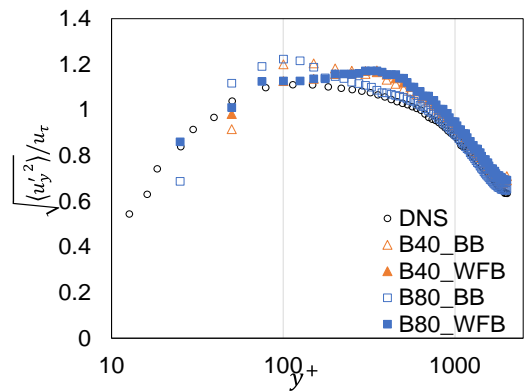

(d) $\sqrt{\langle u_y'^2 \rangle}$, $\mathrm{Re}_\tau = 2003$

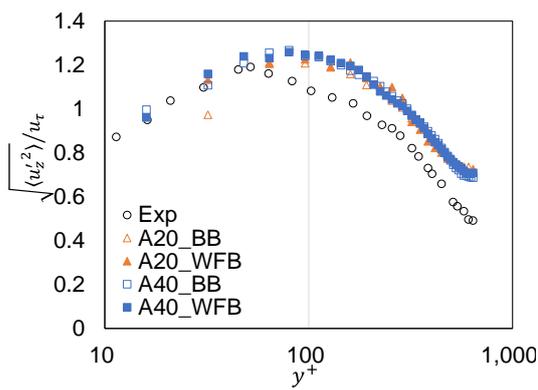

(e) $\sqrt{\langle u_z'^2 \rangle}$, $\mathrm{Re}_\tau = 640$

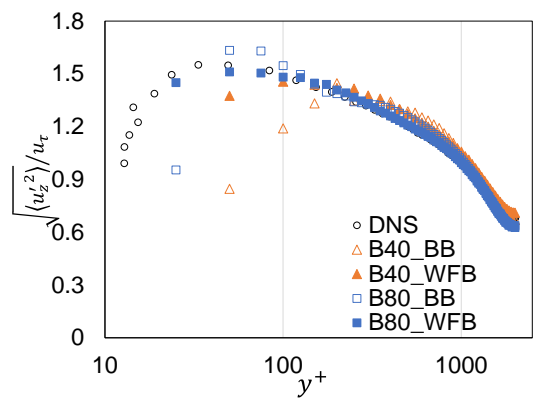

(f) $\sqrt{\langle u_z'^2 \rangle}$, $\mathrm{Re}_\tau = 2003$

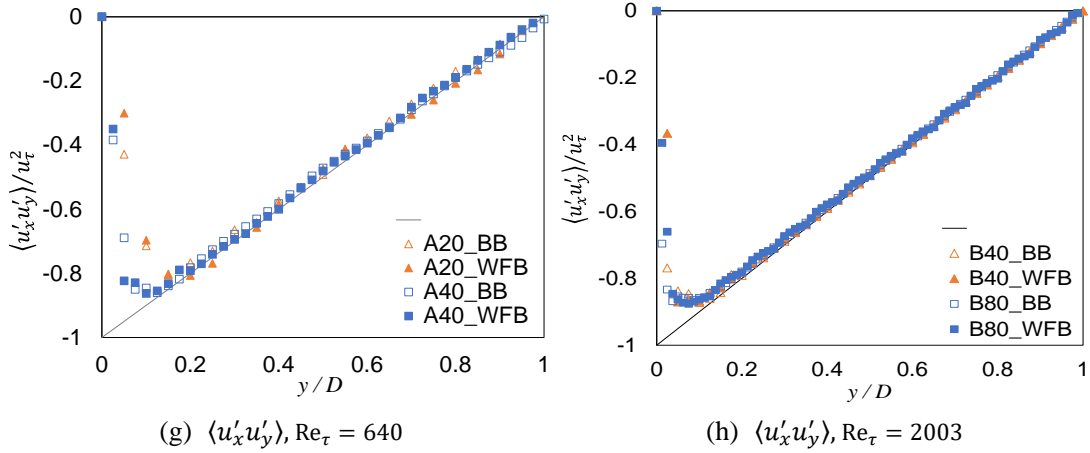

(g) $\langle u'_x u'_y \rangle$, $Re_\tau = 640$
(h) $\langle u'_x u'_y \rangle$, $Re_\tau = 2003$

Fig. 6 Reynolds normal and shear stresses for two different grid resolutions with two different wall boundary conditions in the conditions of $Re_\tau = 640$ (left), and $Re_\tau = 2003$ (right). Experimental data from [46] and [47], and DNS data from [48] were compared.

Fig. 7 depicts the normalized probability distribution functions (PDF) of the coordinate of all the first-layer grids off-wall in terms of inner variables for all cases ($y_1^+$). This is an important attribute of the wall function model because it implies the distribution of the instantaneous shear stress on the wall [52]. The results of FVM-LES with a wall function ($Re_\tau = 2003$) reported by Pantano el al. [52] is also added for reference. Here, $\overline{y_1^+}$ denotes the wall-parallel spatial averages of $y_1^+$. We observed that the PDF of $y_1^+$ in all WFB cases at different $Re_\tau$ scaled well when normalized with its mean and variance, and it also agreed well with Pantano's result. The results, therefore, indicate that the distribution of shear stress reproduced by WFB was in close proximity to that produced by FVM-LES with a wall function model.

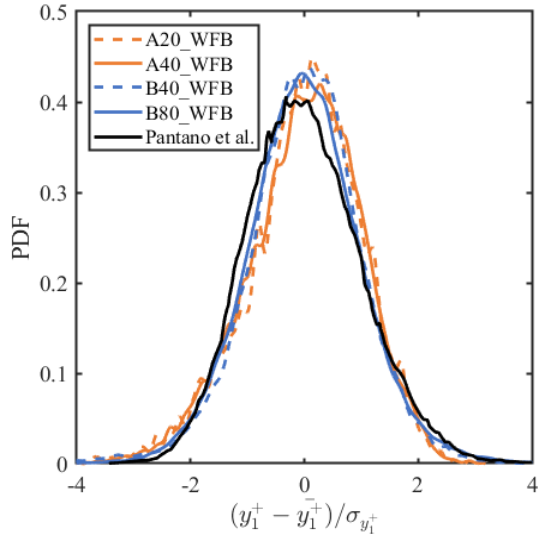

Fig. 7 PDF of the first-layer grids' location $y_1^+$, normalized in terms of $(y_1^+ - \overline{y_1^+})/\sigma_{y_1^+}$.

## 4. Conclusions

In this study, a "WFB" boundary was proposed to incorporate a wall function into the LBM's boundary conditions. Spalding's law was utilized as the wall function. Simulations of a turbulent

Han et al.

channel flow at $\text{Re}_\tau = 640$ and 2003 were implemented using LBM-LES (standard Smagorinsky SGS model) to validate the proposed boundary. The conclusions drawn based on the findings of this study are summarized as follows:

1) The core idea of the WFB is to adjust the distribution functions, which are in the diagonal directions and pointing to the interior of the flow field, to reflect the shear drag obtained from a wall function model.

2) The BB boundary underestimated the time-averaged velocity at the first-layer grids in the buffer layer for both $\text{Re}_\tau = 640$, and 2003. However, the WFB improved it; therefore, the velocity at the first-layer grids agreed well with the Spalding's law. This indicates that the WFB obtained the appropriate shear drag on the wall.

3) The BB boundary underestimated the time-averaged velocity in the entire domain at $\text{Re}_\tau = 2003$ when utilizing coarse grids. The WFB compensated for the underestimation, and the velocity agreed with the experimental data more accurately.

4) The WFB produced similar distributions of the Reynolds normal and shear stresses with the BB, and it improved the Reynolds normal stress in the near-wall region to a certain extent, as compared to the BB.

5) The WFB provided analogous distributions of shear stress on the wall with those of FVM-LES with a wall function.

Therefore, the WFB was established and could partially improve the near-wall accuracy of the LBM-LES in solving the turbulent channel flows compared to the BB boundary. Next, to apply the WFB to the irregular building geometries in wind engineering, we plan to extend it to more complex BB boundaries, such as the interpolated BB boundary. This will be tested and confirmed in our future work.

**Acknowledgments**
This research was partially supported by the Research Fellowships of Japan Society for the Promotion of Science (KAKENHI Grant number: 18J13607, Project general manager: Mengtao Han). Mengtao Han is a JSPS research fellow. This research was also partially supported by the Initiative on Promotion of Supercomputing for Young or Women Researchers, Information Technology Center, The University of Tokyo. The authors also would like to thank Professor Fujihiro Hamba from the Institute of Industrial Science, the University of Tokyo, for his valuable suggestions to this research.

**Appendix A**
D2Q9 and D3Q19 are the widely-utilized schemes to solve 2-dimensional and 3-dimensional flows in DdQq schemes [27]; Fig.A-1 depicts the grid of these schemes.

Han et al.

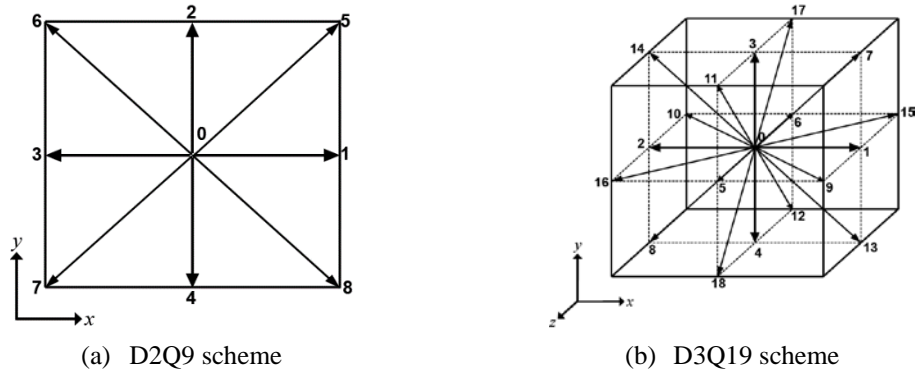

(a) D2Q9 scheme  (b) D3Q19 scheme

Fig.A-1 Grid system of D2Q9 and D3Q19 schemes (reproduced from Krüger[37]).

Fig.A-2 and Fig.A-3 depict the processes of the bounce-back and no-slip boundary in D2Q9 scheme, respectively.

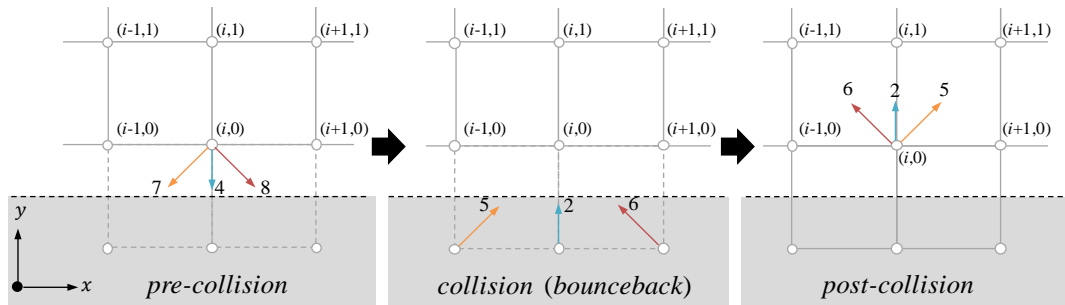

Fig.A-2 Sketch of the BB boundary in D2Q9 scheme (reproduced from Krüger[37]).

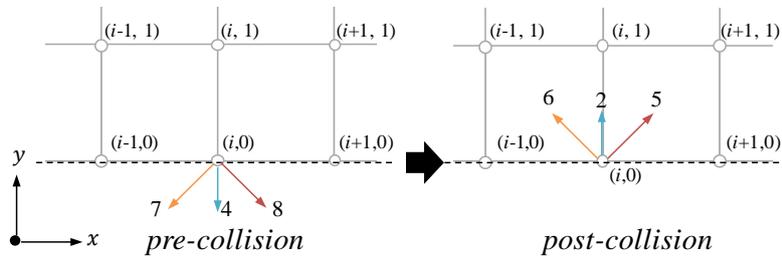

Fig.A-3 Sketch of the free-slip boundary in D2Q9 scheme (reproduced from Krüger[37]).

Table A-1 lists the discrete velocity vectors of D3Q19 scheme [27].

**Table A-1** Discrete velocity vectors $\mathbf{e}_a$ of D3Q19

| $a$ | $\mathbf{e}_a$ | $a$ | $\mathbf{e}_a$ |
|---|---|---|---|
| 0 | (0, 0, 0) | 9, 10 | (±1, 0, ±1) |
| 1, 2 | (±1, 0, 0) | 11, 12 | (0, ±1, ±1) |
| 3, 4 | (0, ±1, 0) | 13, 14 | (±1, ∓1, 0) |
| 5, 6 | (0, 0, ±1) | 15, 16 | (±1, 0, ∓1) |
| 7, 8 | (±1, ±1, 0) | 17, 18 | (0, ±1, ∓1) |

## Appendix B

$\Delta f$ was defined to examine the mass conservation at the wall boundary, which is the average of the differences between $f_a$ before WFB collision and $\widetilde{f_a}^*$ after WFB collision of all boundary grids, as shown in Equation (B-1). Here, $i$ represents the $i^{\text{th}}$ grid at the wall boundary, and $N$ represents the total number of grids on the wall. Furthermore, $\Delta f$ in Table B-1 lists the rate of $\Delta f \neq 0$ in the 5000 steps in all the cases, and the order of the quotient of maximum $\Delta f$ and spatial average $\overline{f_a}$ at the wall boundary.

$$\Delta f = \frac{1}{N}\sum_{i=1}^{N}\left(\sum_{a=0}^{18}\widetilde{f_a}^*|_i - \sum_{a=0}^{18}f_a|_i\right) \tag{B-1}$$

**Table B-1** Probability of $\Delta f \neq 0$ in all cases and the order of maximum $\Delta f$ / $\overline{f_a}$.

| Case name | A20_WFB | A40_WFB | B40_WFB | B40_WFB |
|---|---|---|---|---|
| rate of $\Delta f \neq 0$ | 13.29 % | 13.81 % | 11.35 % | 11.79 % |
| $\dfrac{\text{maximum } \Delta f}{\overline{f_a}}$ | $\sim O(10^{-15})$ | $\sim O(10^{-16})$ | $\sim O(10^{-15})$ | $\sim O(10^{-14})$ |

## References

[1] Dong YH, Sagaut P. A study of time correlations in lattice Boltzmann-based large-eddy simulation of isotropic turbulence. Phys Fluids 2008;20. https://doi.org/10.1063/1.2842381.

[2] Fernandino M, Beronov K, Ytrehus T. Large eddy simulation of turbulent open duct flow using a lattice Boltzmann approach. Math Comput Simul 2009;79:1520–6. https://doi.org/10.1016/j.matcom.2008.07.001.

[3] Han M, Ooka R, Kikumoto H. Lattice Boltzmann method-based large-eddy simulation of indoor isothermal airflow. Int J Heat Mass Transf 2019;130:700–9. https://doi.org/10.1016/j.ijheatmasstransfer.2018.10.137.

[4] Han M, Ooka R, Kikumoto H. Comparison between lattice Boltzmann method and finite volume method for LES in the built environment. 7th Int. Symp. Comput. Wind Eng. 2018, vol. m, Seoul: 2018, p. 2–5.

[5] Sajjadi H, Salmanzadeh M, Ahmadi G, Jafari S. Turbulent indoor airflow simulation using hybrid LES/RANS model utilizing Lattice Boltzmann method. Comput Fluids 2017;150:66–73. https://doi.org/10.1016/j.compfluid.2017.03.028.

[6] Béghein C, Jiang Y, Chen QY. Using large eddy simulation to study particle motions in a room. Indoor Air 2005;15:281–90. https://doi.org/10.1111/j.1600-0668.2005.00373.x.

[7] Inamuro T. The Lattice Boltzmann Method and Its Applications for Complex Flows. J Soc Powder Technol Japan 1999;36:286–91. https://doi.org/10.4164/sptj.36.286.

[8] WU C-J, GUAN H. Lattice Boltzmann Dynamics and Dynamical System Sub-Grid Models. Mod Phys Lett B 2009;23:349–52. https://doi.org/10.1142/S0217984909018370.

[9] Zhuo C, Zhong C. LES-based filter-matrix lattice Boltzmann model for simulating turbulent natural convection in a square cavity. Int J Heat Fluid Flow 2013;42:10–22. https://doi.org/10.1016/j.ijheatfluidflow.2013.03.013.

[10]   Kuwata Y, Suga K. Large eddy simulations of pore-scale turbulent flows in porous media by the lattice Boltzmann method. Int J Heat Fluid Flow 2015;55:143–57. https://doi.org/10.1016/J.IJHEATFLUIDFLOW.2015.05.015.

[11]   Zhou X, Dong B, Chen C, Li W. A thermal LBM-LES model in body-fitted coordinates: Flow and heat transfer around a circular cylinder in a wide Reynolds number range. Int J Heat Fluid Flow 2019;77:113–21. https://doi.org/10.1016/J.IJHEATFLUIDFLOW.2019.04.001.

[12]   Suga K, Chikasue R, Kuwata Y. Modelling turbulent and dispersion heat fluxes in turbulent porous medium flows using the resolved LES data. Int J Heat Fluid Flow 2017;68:225–36. https://doi.org/10.1016/J.IJHEATFLUIDFLOW.2017.08.005.

[13]   Versteeg, H.K. and Malalasekera W. An introduction to computational fluid dynamics: the finite volume method. Pearson Education. 2007;44.

[14]   Grötzbach G. Direct numerical and large eddy simulations of turbulent channel flows. Encycl Fluid Mech 1987;6:1337–91.

[15]   Werner H, Wengle H. Large-Eddy Simulation of Turbulent Flow Over and Around a Cube in a Plate Channel. Turbul. Shear Flows 8, Berlin, Heidelberg: Springer Berlin Heidelberg; 1993, p. 155–68. https://doi.org/10.1007/978-3-642-77674-8_12.

[16]   Launder BE, Spalding DB. The numerical computation of turbulent flows. Comput Methods Appl Mech Eng 1974;3:269–89. https://doi.org/10.1016/0045-7825(74)90029-2.

[17]   Spalding DB. A Single Formula for the "Law of the Wall." J Appl Mech 1961;28:455. https://doi.org/10.1115/1.3641728.

[18]   Stathopoulos T, Baskaran BA. Computer simulation of wind environmental conditions around buildings. Eng Struct 1996;18:876–85. https://doi.org/10.1016/0141-0296(95)00155-7.

[19]   Toparlar Y, Blocken B, Vos P, van Heijst GJF, Janssen WD, van Hooff T, et al. CFD simulation and validation of urban microclimate: A case study for Bergpolder Zuid, Rotterdam. Build Environ 2015;83:79–90. https://doi.org/10.1016/J.BUILDENV.2014.08.004.

[20]   Kikumoto H, Ooka R, Han M, Nakajima K. Consistency of mean wind speed in pedestrian wind environment analyses: Mathematical consideration and a case study using large-eddy simulation. J Wind Eng Ind Aerodyn 2018;173:91–9. https://doi.org/10.1016/j.jweia.2017.11.021.

[21]   Cornubert R, d'Humières D, Levermore D. A Knudsen layer theory for lattice gases. Phys D Nonlinear Phenom 1991;47:241–59. https://doi.org/10.1016/0167-2789(91)90295-K.

[22]   Ziegler DP. Boundary conditions for lattice Boltzmann simulations. J Stat Phys 1993;71:1171–7. https://doi.org/10.1007/BF01049965.

[23]   Liu X, Guo Z. A lattice Boltzmann study of gas flows in a long micro-channel. Comput Math with Appl 2013;65:186–93. https://doi.org/10.1016/j.camwa.2011.01.035.

[24]   Geller S, Krafczyk M, Tölke J, Turek S, Hron J. Benchmark computations based on lattice-Boltzmann, finite element and finite volume methods for laminar flows. Comput Fluids 2006;35:888–97. https://doi.org/10.1016/j.compfluid.2005.08.009.

[25]   Fakhari A, Lee T. Numerics of the lattice boltzmann method on nonuniform grids: Standard LBM and finite-difference LBM. Comput Fluids 2015;107:205–13. https://doi.org/10.1016/j.compfluid.2014.11.013.

[26]   Breuer M, Bernsdorf J, Zeiser T, Durst F. Accurate computations of the laminar flow past a square cylinder based on two different methods: lattice-Boltzmann and finite-volume. Int J Heat Fluid Flow 2000;21:186–96. https://doi.org/10.1016/S0142-727X(99)00081-8.

[27] Qian YH, D'Humières D, Lallemand P. Lattice bgk models for navier-stokes equation. Epl 1992;17:479–84. https://doi.org/10.1209/0295-5075/17/6/001.

[28] Lagrava D, Malaspinas O, Latt J, Chopard B. Advances in multi-domain lattice Boltzmann grid refinement. J Comput Phys 2012;231:4808–22. https://doi.org/10.1016/j.jcp.2012.03.015.

[29] Gendre F, Ricot D, Fritz G, Sagaut P. Grid refinement for aeroacoustics in the lattice Boltzmann method: A directional splitting approach. Phys Rev E 2017;96:023311. https://doi.org/10.1103/PhysRevE.96.023311.

[30] Li W, Luo LS. Finite Volume Lattice Boltzmann Method for Nearly Incompressible Flows on Arbitrary Unstructured Meshes. Commun Comput Phys 2016;20:301–24. https://doi.org/10.4208/cicp.211015.040316a.

[31] Geier M, Schönherr M, Pasquali A, Krafczyk M. The cumulant lattice Boltzmann equation in three dimensions: Theory and validation. Comput Math with Appl 2015;70:507–47. https://doi.org/10.1016/j.camwa.2015.05.001.

[32] Gehrke M, Janßen CF, Rung T. Scrutinizing lattice Boltzmann methods for direct numerical simulations of turbulent channel flows. Comput Fluids 2017;156:247–63. https://doi.org/10.1016/j.compfluid.2017.07.005.

[33] Kawai S, Larsson J. Wall-modeling in large eddy simulation: Length scales, grid resolution, and accuracy. Phys Fluids 2012;24:015105. https://doi.org/10.1063/1.3678331.

[34] Norouzi A, Esfahani JA. Two relaxation time lattice Boltzmann equation for high Knudsen number flows using wall function approach. Microfluid Nanofluidics 2014;18:323–32. https://doi.org/10.1007/s10404-014-1435-6.

[35] Malaspinas O, Sagaut P. Wall model for large-eddy simulation based on the lattice Boltzmann method. J Comput Phys 2014;275:25–40. https://doi.org/10.1016/j.jcp.2014.06.020.

[36] Pasquali A, Geier M, Krafczyk M. Near-wall treatment for the simulation of turbulent flow by the cumulant lattice Boltzmann method. Comput Math with Appl 2020;79:195–212. https://doi.org/10.1016/j.camwa.2017.11.022.

[37] Krüger T, Kusumaatmaja H, Kuzmin A, Shardt O, Silva G, Viggen EM. The Lattice Boltzmann Method: Principles and Practice. Springer; 2017. https://doi.org/10.1007/978-3-319-44649-3.

[38] Bouzidi M, Firdaouss M, Lallemand P. Momentum transfer of a Boltzmann-lattice fluid with boundaries. Phys Fluids 2001;13:3452–9. https://doi.org/10.1063/1.1399290.

[39] Moin P, Kim J. Numerical investigation of turbulent channel flow. J Fluid Mech 1982;118:341–77. https://doi.org/10.1017/S0022112082001116.

[40] d'Humieres D, Ginzburg I, Krafczyk M, Lallemand P, Luo L-S. Multiple-relaxation-time lattice Boltzmann models in three dimensions. Philos Trans R Soc A Math Phys Eng Sci 2002;360:437–51. https://doi.org/10.1098/rsta.2001.0955.

[41] Rogallo RS, Moin P. Numerical Simulation of Turbulent Flows. Annu Rev Fluid Mech 1984;16:99–137. https://doi.org/10.1146/annurev.fl.16.010184.000531.

[42] Driest ER Van. On Turbulent Flow Near a Wall. J Aeronaut Sci 1956;23:1007–11. https://doi.org/10.2514/8.3713.

[43] Klainerman S, Majda A. Compressible and incompressible fluids. Commun Pure Appl Math 1982;35:629–51. https://doi.org/10.1002/cpa.3160350503.

[44] Martínez DO, Matthaeus WH, Chen S, Montgomery DC. Comparison of spectral method and lattice Boltzmann simulations of two-dimensional hydrodynamics. Phys Fluids 1994;6:1285–

98. https://doi.org/10.1063/1.868296.

[45] Reider MB, Sterling JD. Accuracy of discrete-velocity BGK models for the simulation of the incompressible Navier-Stokes equations. Comput Fluids 1995;24:459–67. https://doi.org/10.1016/0045-7930(94)00037-Y.

[46] Hussain AKMF, Reynolds WC. Measurements in Fully Developed Turbulent Channel Flow. J Fluids Eng 1975;97:568. https://doi.org/10.1115/1.3448125.

[47] Clark JA. A Study of Incompressible Turbulent Boundary Layers in Channel Flow. J Basic Eng 1968;90:455. https://doi.org/10.1115/1.3605163.

[48] Hoyas S, Jiménez J. Scaling of the velocity fluctuations in turbulent channels up to Reτ=2003. Phys Fluids 2006;18:10–4. https://doi.org/10.1063/1.2162185.

[49] Ferziger JH, Perić M. Computational Methods for Fluid Dynamics. vol. 46. Berlin, Heidelberg: Springer Berlin Heidelberg; 2002. https://doi.org/10.1007/978-3-642-56026-2.

[50] Mathew J. Large eddy simulation. Def Sci J 2010;60:598–605. https://doi.org/10.14429/dsj.60.602.

[51] Piomelli U, Ferziger JH, Moin P. Models for large eddy simulations of turbulent channel flows including transpiration. Stanford Univ Rep 1987:TF-32.

[52] Pantano C, Pullin DI, Dimotakis PE, Matheou G. LES approach for high Reynolds number wall-bounded flows with application to turbulent channel flow. J Comput Phys 2008;227:9271–91. https://doi.org/doi.org/10.3130/aija.62.47_1.